\documentstyle[preprint,aps,eqsecnum]{revtex}
\tighten
\begin{document}
\preprint{PITT-96-172/ IFP-605-UNC}
\draft
\title{{\bf   RELAXATION AND KINETICS \\
        IN SCALAR FIELD THEORIES}}
\author{{\bf D. Boyanovsky$^{(a)}$,
 I. D. Lawrie$^{(b)}$ and
 D.-S. Lee$^{(c)}$}}
\address
{ (a)  Department of Physics and Astronomy, University of
Pittsburgh, Pittsburgh, PA. 15260, U.S.A. \\
(b) Department of Physics and Astronomy, University of Leeds,
Leeds LS2 9JT, U.K.\\
(c) Department of Physics and Astronomy, University of North
Carolina\\
Chapel Hill, N.C. 27599, U.S.A. }
\date{ }
\maketitle
\begin{abstract}
A new approach to the dynamics of relaxation and kinetics of
thermalization in a scalar field theory is presented that incorporates
the relevant time scales through the resummation of hard thermal
loops. An alternative derivation of the kinetic equations for the
``quasiparticle'' distribution functions is obtained that allows a
clear understanding of the different ``coarse graining''
approximations usually involved in a kinetic description. This method
leads to a systematic perturbative expansion to obtain the kinetic
equations including hard-thermal loop resummation and to an
improvement including renormalization, off-shell effects and
contributions that change chemical equilibrium on short time scales.
As a byproduct of these methods we establish the relation between the
relaxation time scale in the linearized equation of motion of the
quasiparticles and the thermalization time scale of the quasiparticle
distribution function in the ``relaxation time approximation''. Hard
thermal loop resummation dramatically modifies the scattering rate for
long wavelength modes as compared to the usual (semi) classical
estimate. Relaxation and kinetics are studied both in the unbroken and
broken symmetry phases of the theory. The broken symmetry phase also
provides the setting to obtain the contribution to the kinetic
equations from processes that involve decay of a heavy scalar into
light scalar particles in the medium.
\end{abstract}

\section{\bf Introduction}
The motivation for understanding non-equilibrium phenomena in
intermediate and high energy physics becomes more pressing with the
experimental possibility of probing the quark-gluon and chiral phase
transitions in upcoming heavy ion colliders at RHIC and large
energy-luminosity hadron colliders at LHC. In these high energy, high
luminosity experiments ($\sqrt{s} \geq 200 \mbox{Gev}/\mbox{nucleon}$)
a large energy density $\approx 1 \mbox{Gev}/ \mbox{fm}^3$ is
deposited in the collision region corresponding to temperatures $ T
\approx 200 \mbox{Mev}$ with time scales for the relevant relaxational
processes of order $t \approx 1\mbox{fm}/c$. In these extreme
situations, the challenge is to describe transport and relaxation
phenomena from a fundamental microscopic theory since a description of
non-equilibrium processes may lead to a better understanding of the
experimental signatures of the transition and evolution of the
plasma\cite{muller1}. Any model that attempts to describe the
formation and evolution of the quark-gluon plasma and eventual
hadronization in heavy ion collisions will have to succeed in
providing a description of thermalization and relaxation during the
expansion stage. The rate of thermalization and relaxation will
determine if an equilibrium thermodynamical description is suitable or
a full non-equilibrium treatment will ultimately be necessary to make
quantitative statements on the evolution and signatures of the phase
transition. Thermalization and relaxation are usually studied via a
kinetic approach to non-equilibrium phenomena, typically through
Boltzmann equations for the distribution functions. Such a description
involves a wide separation of time and length scales and uncontrolled
and drastic approximations are often made to simplify the problem.
There is considerable effort in obtaining non-equilibrium equations
that describe thermalization and relaxation starting from the
microscopic theory that describes the full dynamics
\cite{daniel1,elze,hu,henning,rau,heinz}. More recently this program
has been extended to a detailed study of the kinetics of QCD and the
excitations of non-abelian gauge theories \cite{geiger,weldon1,iancu}
and the transport coefficients in scalar theories \cite{jeon}.

Transport equations for quark matter are the subject of intense study
since a deeper understanding of the thermalization and hadronic
processes needs a full microscopic description. A kinetic description
based on QCD has been attempted over a decade ago but remains a
formidable formal structure with very difficult implementation. While
a description from a fundamental theory is desirable, the technical
obstacles are fairly formidable and instead a description in terms of
low energy effective theories is emerging \cite{zhang,mrow,klebansky}.

A kinetic description of transport and relaxation is justified only
when there is a wide separation between the microscopic time and
length scales, namely the thermal (or Compton) wavelength (mean
separation of particles) and the relaxation scales (mean free path and
relaxation time). The usual kinetic approach leading to Boltzmann
transport equations involves Wigner transforms of two point (or higher
point) correlation functions and eventually a gradient expansion
\cite{henning,zhang,mrow}. Such a gradient expansion assumes that the
center of mass Wigner variables are ``slowly varying'' but it is
seldom clear at this level which are the fast and which are the slow
scales involved. A ``coarse graining'' procedure is typically invoked
that averages out microscopic scales in the kinetic description
\cite{hu} leading to irreversible evolution in the resulting
equations. Such averaging procedure is usually poorly understood and
justified {\it a posteriori}.

It is recognized that at temperatures much larger than the masses of
the particles, it is necessary to perform a resummation of the
perturbative series to incorporate the relevant microscopic time
scales in the description of thermal processes. Such a
non-perturbative resummation scheme has been proposed by Braaten and
Pisarski \cite{pisarski1} and since then used to obtain damping rates
\cite{iancu,blaizot}. A program to incorporate self-consistently the
microscopic time scales within a kinetic approach has been proposed by
Lawrie \cite{lawrie} in a manner that is similar in some respects to
the hard-thermal loop resummation. In the quark-gluon plasma this
program acquires further significance, since for temperatures $T
\approx 200 \mbox{Mev}$ the thermal masses of the ``light quarks'' are
of order $M_q(T) \approx gT > m_q \approx 5-10 \rm{Mev}$ with $m_q$
the ``current'' quark masses and $g$ the quark-gluon coupling constant
\cite{weldon1,iancu,pisarski1}.

Another realm in which relaxation and thermalization also play a
fundamental role is the description of the reheating stage in
cosmological inflationary scenarios \cite{revs,kolb,lindebook}.
Recently new non-equilibrium phenomena have been recognized to play a
role during the reheating stage in the post-inflationary epoch in
early universe cosmology. These phenomena are the result of profuse
particle production during the stage of parametric amplification of
quantum fluctuations for large amplitudes of the inflaton field
\cite{kofman,branden3,reheating,yoshimura}. A similar mechanism has
been recently proposed to play a role in hadronization during the
supercooling stage of the quark-gluon phase transition \cite{muller2}.
Since the particles produced during this stage are in a
non-equilibrium distribution \cite{reheating}, the final stage of
reheating needs a deeper understanding of thermalization processes.

The reliability of any kinetic approach to the description of
thermalization and relaxation, either in early universe processes or
hadronization and equilibration during the quark-gluon plasma
transition, hinges upon a {\em complete} understanding of the time
scales involved. Since in extreme environments the microscopic time
scales are modified by the medium, it is important to incorporate
these effects in the kinetic description.

The purpose of this article is to provide an alternative program for
the description of relaxation and kinetics in scalar field theories.
The relevant microscopic and relaxational time scales are incorporated
in a consistent first principles calculation starting from the
microscopic theory. Such an approach allows us to identify the several
different ``coarse graining'' procedures with a detailed understanding
of the relevant scales. The resummation of the hard thermal loops is
incorporated consistently in the derivation of the equations of motion
that describe relaxation and kinetics and provides a systematic scheme
to include off-shell and higher order corrections to the resulting
kinetic equations, including terms that provide departure from
chemical equilibrium on short time scales, and renormalization
effects. We find that the hard thermal loop resummation modifies
dramatically the estimate of the scattering rate in the Boltzmann
equation for long wavelength modes, from that of a ``naive'' or
(semi)classical argument.

Furthermore, our approach provides a direct proof of the relation
between the relaxation processes of quasi-particles and the
thermalization rate of the quasiparticle distribution function in the
``relaxation time approximation'' of the Boltzmann equation that
describes their evolution. Such a relation was proposed by Weldon for
the identification of the quasi-particle damping rate \cite{weldon}
but to our knowledge a complete proof of such relation has not been
given within the context of the Boltzmann equation as derived from a
microscopic theory.

Although we focus here on a scalar field theory both in the broken and
unbroken symmetry phase to distinguish the different scales and
processes, the methods can be extended to other models and theories
almost straightforwardly (obviously with the complications of gauge
invariance in gauge theories). The broken symmetry case allows us to
study the situation in which the scalar field is coupled to other
lighter scalars with a trilinear coupling and to include decay and
recombination processes in both the relaxation and kinetic
description.

In the next section we introduce the model and the techniques. In
section III we study the relaxation of ``quasi-particles'' obtaining
the real-time equation of motion in the linearized approximation,
including consistently the hard-thermal-loop corrections to the
self-energy. In section IV we introduce a new formulation of kinetic
theory that incorporates the hard-thermal-loop corrections
consistently, displays all the relevant time scales and leads to a
well defined ``coarse-graining'' procedure. This formulation allows a
systematic improvement on the kinetic description and, if necessary,
consistently includes off-shell effects and renormalization in a
computationally accesible form.

Section V presents our conclusions, summarizes the main results and
poses new questions.

\section{\bf The Model and Techniques}
We will restrict our study to a self-interacting scalar theory both in
the broken and unbroken symmetry state. The Lagrangian density is
given by:
\begin{equation}
{\cal{L}}_0 = {1 \over 2} \left( \partial_{\mu} \Phi \right) ^2 - {1
\over 2} m^2_{0} \Phi^2 -{ \lambda_0 \over 4!} \Phi^4 \;.
                                                  \label{lagrangian}
\end{equation}
with $m_0\; ; \lambda_0$ the bare parameters.

As mentioned in the introduction, the first step towards understanding
the kinetic regime is the identification of the {\em microscopic} time
scales in the problem. In a medium, the bare particles are dressed by
the interactions becoming ``quasiparticles''. One is interested in
describing the relaxation of these quasiparticles. Thus the important
microscopic time scales are those associated with the quasiparticles
and not the bare particles. If a Boltzmann equation is obtained in
some perturbative scheme, such a scheme should be in terms of the
quasiparticles, which already implies a resummation of the
perturbative expansion. This is precisely the rationale behind the
resummation of the hard thermal loops in finite temperature field
theory \cite{pisarski1} and also behind the self-consistent treatment
proposed by Lawrie \cite{lawrie}. In this scalar theory such a
resummation can be implemented by writing in the Lagrangian
\begin{equation}
m^2_0= m^2_R(T)+\delta m^2(T)                      \label{counterterm}
\end{equation}
where $m^2_R(T)$ is the renormalized and {\em temperature dependent}
quasiparticle mass which enters in the propagators and $\delta m^2(T)$
is a counterterm which will cancel a subset of Feynman diagrams in the
perturbative expansion and is considered part of the interaction
Lagrangian. This method is the simplest form of implementing the hard
thermal loop resummation in the scalar case and was used within this
context by Parwani \cite{parwani}. We note that this renormalized,
temperature dependent mass determines the important time scales in the
medium but is {\em not} the position of the quasiparticle pole (or,
stricly speaking, resonance: see the discussion below). We could also
introduce counterterms for wave-function and coupling constant
renormalization and proceed to a perturbative expansion of the BPHZ
type in terms of the renormalized couplings and fields. Such a
possibility will be discussed later within the context of improvements
on the method and renormalization.

\subsection{Non- Equilibrium Techniques}

The field theoretical methods to describe processes out of equilibrium
are known and described at length in the literature
\cite{S,K,maha,CSHY,kapusta}. The basic ingredient is the time
evolution of an initially prepared density matrix, which leads to the
generating functional of non-equilibrium Green's functions in terms of
a path integral representation along a contour in the complex time
plane. This contour involves a forward time branch, a backward time
branch and a third branch down the imaginary time axis to time $\tau =
-i \beta$ if the initial density matrix describes an equilibrium
ensemble at initial temperature $1/\beta$. For the computation of
real-time correlation functions at the order of approximation we
consider, the only role of the imaginary time branch is to determine
the boundary conditions on the propagators.

The fields living on the forward and backward branches will be
labelled with $+$ and $-$, respectively, and the effective Lagrangian
that enters in the path integral representation of the non-equilibrium
generating functional is given by
\begin{equation}
{\cal{L}}_{\rm{noneq}}=
{\cal{L}}[{\Phi}^+]-{\cal L}[{\Phi}^-]               \label{noneqlag}.
\end{equation}

{F}rom this path integral representation it is possible to construct
a perturbative expansion of the non-equilibrium Green's functions in
terms of modified Feynman rules. Correlation functions are obtained as
functional derivatives with respect to sources $j^{\pm}$ on the
respective branches. The non-equilibrium Feynman rules are:

i) The number of vertices is doubled: those in which all the fields
are on the $+$ branch are the usual interaction vertices, while those
in which the fields are on the $-$ branch have the opposite sign.

ii) The combinatoric factors are the same as in usual field theory.

iii) The spatial Fourier transform of the (bosonic) propagators are
\begin{eqnarray}
G_k^{++}(t,t')&=&G_k^{>}(t,t')\Theta(t-t')+G_k^{<}(t,t')
\Theta(t'-t)\; ,                                   \label{gplusplus}\\
G_k^{--}(t,t')&=&G_k^{>}(t,t')\Theta(t'-t)+G_k^{<}(t,t')
\Theta(t-t')\;,                                  \label{gminusminus}\\
G_k^{+-}(t,t')&=&- G_k^{<}(t,t') \;,             \label{gplusminus} \\
G_k^{-+}(t,t')&=&- G_k^{>}(t,t') \;,             \label{gminusplus} \\
G_k^{>}(t,t')&=&i \int d^3x  \; e^{-i\vec{k}\cdot\vec{x}} \;
\langle \Phi(\vec{x},t) \Phi(\vec{0},t') \rangle \;, \label{greater}\\
G_k^{<}(t,t')&=&i \int d^3x \; e^{-i\vec{k}\cdot\vec{x}} \;
\langle \Phi(\vec{0},t') \Phi(\vec{x},t) \rangle \;.  \label{gsmaller}
\end{eqnarray}
where $\Phi$ is the bose field. Now we have to specify the properties
of the initial state. Particularly convenient is the choice of a
thermal initial state with temperature $T$. Then the density matrix of
this initial state is $\rho=e^{-H_0/T}$, where $H_0$ is the
Hamiltonian for times $t<0$. This choice of the initial state
determines the boundary conditions on the Green's functions. These are
the usual periodicity conditions in imaginary time (KMS conditions):
\begin{equation}
G^<(\vec x,t;\vec x',t')=G^>(\vec x,t-i\beta;\vec x',t')   \label{kms}
\end{equation}

Finally, the free-field Green's functions for bose fields are
constructed from the following ingredients:
\begin{eqnarray}
G_k^{>}(t,t')&=&\frac{i}{2\omega_k}\left\{[1+n(\omega_k)]
e^{-i\omega_k(t-t')}+n(\omega_k) e^{i\omega_k(t- t')} \right\}\;,
                                                      \label{ggreat}\\
G_k^{<}(t,t')&=&\frac{i}{2\omega_k}\left\{[1+n(\omega_k)]
e^{i\omega_k(t-t')}+n(\omega_k)e^{-i\omega_k(t- t')}\right\}\;,
                                                      \label{gsmall}\\
\omega_k&=&\sqrt{\vec{k}^2+m^2_R(T)}\;,\quad\quad\quad
n_{b}(\omega_k)=\frac{1}{e^{\beta \omega_k}-1} \;,
                                                    \label{bosefactor}
\end{eqnarray}
An important property that will be used in the calculations that
follow is the relation
\begin{equation}
G_k^{>}(t,t') = G_k^{<}(t',t) \label{relation}
\end{equation}

\section{\bf Linear Relaxation}

In this section we study the relaxation of quasiparticles, from the
linearized real-time equation of motion for the {\em expectation
value} of the scalar field
\begin{equation}
\phi(\vec{x},t)\equiv \langle \Phi^{+} (\vec{x},t) \rangle =\langle
\Phi^{-} (\vec{x},t) \rangle
                                                    \label{expecvalue}
\end{equation}
where $\Phi^{\pm}$ are the fields on the forward and backward branches
of the contour respectively and the expectation value is in the
ensemble of the initial density matrix. We derive the equation of
motion for $\phi(\vec x, t)$ using the tadpole method \cite{B3,dslee},
which consists in splitting the field $\Phi$ into its expectation
value and a fluctuation as
\begin{equation}
\Phi^{\pm}(\vec{x},t)=\phi(\vec{x},t)+\psi^{\pm}(\vec{x},t)\;,
                                                         \label{shift}
\end{equation}
where $\psi^{\pm}$ are the fluctuation operators defined along the
respective branches. The effective equation of motion for the field
$\phi(\vec{x},t)$ follows from the condition:
\begin{equation}
<\psi^{\pm}(\vec{x},t)>=0\;.                               \label{tad}
\end{equation}

This tadpole method is different from that used by Elmfors {\it et al}
\cite{elmfors} in that the present method is implemented with the
non-equilibrium, real time functional and Green's functions.

After the shift (\ref{shift}), the non-equilibrium Lagrangian density
reads

\begin{eqnarray}
{\cal{L}}[\phi+\psi^+]-{\cal{L}}[\phi+\psi^-]&=&\left\{ \frac{\delta
{\cal{L}}}{\delta \phi}\psi^+ + {\cal{L}}_0[\psi^+] -
\lambda\left(\frac{\phi^2(\psi^+)^2}{4}+
\frac{\phi(\psi^+)^3}{6}+\frac{(\psi^+)^4}{4!}\right) \right.
\nonumber \\
&&\qquad \left. -\frac{1}{2}\delta m^2(T)(2\phi \psi^+ +
(\psi^+)^2)\right\}- \left\{\psi^+ \rightarrow \psi^-\right\}
                                                  \label{plusminuslag}
\end{eqnarray}
with ${\cal{L}}_0[\psi^{\pm}]$ the free field Lagrangian density of a
field with mass $m_R(T)$. To implement the tadpole method one
considers the {\em linear}, cubic and quartic terms and counterterm as
perturbations and the condition (\ref{tad}) is imposed order by order
in a perturbative expansion, using the Feynman rules given
above\cite{B3,dslee}.

We now analyze different cases in turn.

\subsection{\bf $T>>T_c$: Unbroken symmetry case}
We are interested in studying relaxation and kinetics both in the
disordered high temperature phase as well as in the ordered low
temperature phase. This is achieved by allowing the zero temperature
renormalized squared mass to be negative, that is
\begin{equation}
m^2_R(T=0) = -|m^2_R(T=0)| .                           \label{negmass}
\end{equation}
When the temperature is much larger than the renormalized (zero
temperature) mass, the hard-thermal loop resummation is needed to
incorporate the physically relevant time and length scales in the
perturbative expansion. In the equation of motion this is achieved by
requiring that the mass counterterm $\delta m^2(T)$ cancel the 1-loop
tadpole contribution.

This leads to the following self-consistent ``gap'' equation for
$m^2_R(T)$ \cite{parwani,dolan}:
\begin{equation}
m^2_{R} (T) = -m^2_0 + {\lambda \over 2} \int { dk \over 2\pi^2}
{k^2 \over 2 \sqrt{ k^2+m^2_R(T) }} \left\{ 1+ { 2 \over e^{\beta
\sqrt{ k^2+m^2_{R}(T) }}-1} \right\}                     \label{gapeq}
\end{equation}

The divergence in the zero temperature part of (\ref{gapeq})
(quadratic and logarithmic in terms of a spatial momentum cutoff) can
be absorbed in a renormalization of the bare mass by a subtraction at
some renormalization scale:
\begin{equation}
m^2_{R} (T) =-|m^2_R| + {\lambda \over 2} \int { dk \over 2\pi^2} {k^2
\over 2 \sqrt{ k^2+m^2_R(T) }} \left\{ 1+ { 2 \over e^{\beta \sqrt{
k^2+m^2_{R}(T) }}-1} \right\}- \mbox{subtraction}     \label{gapeqren}
\end{equation}
where both $m^2_R$ and  the $\mbox{subtraction}$ depend on the
renormalization scale and the renormalization prescription. Now the
gap equation is finite. For $T \gg m_R(T)$ we obtain

\begin{eqnarray}
m^2_{R}(T)&=&-|m^2_R|+{\lambda \over 2}\left\{\ {T^2 \over 12 }-
{m_{R}(T) T \over 4 \pi} + {\cal{O}} \left(m_{R}(T) \log {m^2_{R}(T)
\over T^2}\right) \right\}
\; \;\quad {\rm{for}} \quad T > m_R(T) \nonumber \\
&=&{\lambda T^2 \over 24} + {\cal{O}}\left(\lambda^{3
\over 2} T^2\right) \; \quad \; \;\quad {\rm{for}} \quad T >
{\sqrt{\lambda}} T >|m_R|                               \label{sgapeq}
\end{eqnarray}
The leading term of the last expression of (\ref{sgapeq}) provides the
correct microscopic time scale. In the massless case it serves as an
infrared cutoff for the loop integrals \cite{parwani,elmfors}. We will
concentrate on the {\em linear relaxation} case in which the equations
of motion for the expectation value are linearized in the amplitude of
$\phi(\vec x, t)$. This approximation is valid for small amplitude
fluctuations from the minimum of the finite temperature effective
potential in the unbroken symmetry phase. Two-loop diagrams
contributing to the equation of motion obtained from the condition
$\langle \psi^+\rangle = 0$ are shown in figure 1. This condition
leads to
\begin{equation}
\ddot{\phi}(\vec{x}',t')-\nabla^2 \phi(\vec{x}',t') +m^2_{R}(T)
\phi(\vec{x}',t') +\int d^3x'' dt''
\Sigma_{ret}(\vec{x'}-\vec{x''},t'-t'') \phi(\vec{x}'',t'') =0
                                                      \label{eqnofmot}
\end{equation}
The mass counterterm has been used to cancel the 1-loop tadpoles. To
two-loop order we find
\begin{equation}
\Sigma_{ret} (\vec{x}-\vec{x'},t-t') = -i \frac{\lambda^2}{6}
[(-iG^>(\vec{x}-\vec{x'},t-t'))^3 - (-iG^< (\vec{x}-\vec{x'},t-t'))^3]
\Theta(t-t')                                          \label{sigmaret}
\end{equation}
The equation obtained from $\langle \psi^- \rangle =0$ is the same as
(\ref{eqnofmot}) as a consequence of unitarity.

Translational invariance of the self-energy makes it convenient to
write the equation of motion for the spatial Fourier transform of
$\phi$. We therefore introduce
\begin{eqnarray}
\phi({\vec x},t) & = & \int{{d^3 p}\over{(2\pi)^3}} \;e^{i \vec p
\cdot \vec x}\,\phi_{\vec p}(t) \label{invF} \\
\Sigma_{ret}(\vec{x}-\vec{x},t-t') & = & \int{{d^3 p}\over{(2\pi)^3}}
\;e^{i \vec p \cdot (\vec x - \vec{x}')}\,\Sigma_{\vec p}(t-t')
                                                        \label{sigmaF}
\end{eqnarray}

In order to solve the integro-differential equation, we will impose
the initial condition $\dot{\phi}_{\vec p}(t<0)=0$ such that this
configuration is ``released'' at time $t=0$ \cite{B3}. Under this
condition the evolution equation becomes
\begin{equation}
\ddot{\phi}_{\vec p}(t)+\omega^2_p \phi_{\vec p}(t)+\int_0^t dt'
\Sigma_{\vec p}( t-t')\phi_{\vec p}(t')=0         \label{finaleqofmot}
\end{equation}
where
\begin{equation}
\omega^2_p = |\vec{p}|^2+m^2_R(T)\\                  \label{sigmafreq}
\end{equation}
\begin{eqnarray}
\Sigma_{\vec{p}}(t-t')&=&-\frac{\lambda^2}{3}\int {d^3 k_1 \over
(2 \pi)^3} \; {d^3 k_2 \over (2 \pi)^3} \; {d^3 k_3 \over (2 \pi)^3}
\; {1 \over 8 \omega_{k_1} \omega_{k_2} \omega_{k_3} }
\left[ (2 \pi)^3 \delta^3 ({\vec{p}}- {\vec{k}}_1 - {\vec{k}}_2-
{\vec{k}}_3) \right] \nonumber \\
&& \left\{\sin \left[(\omega_{k_1} +\omega_{k_2} +\omega_{k_3} )
(t -t') \right]
\left[(1+n_{k_1})\,(1+n_{k_2})\,(1+n_{k_3}) -\; n_{k_1}\, n_{k_2}\,
n_{k_3}\,\right] \right. \nonumber \\
&&+ \left. 3 \sin \left[ ( \omega_{k_1} +\omega_{k_2} -\omega_{k_3})
(t -t') \right] \left[(1+n_{k_1})\, (1+n_{k_2})\,n_{k_3} -\; n_{k_1}
\, n_{k_2} \, (1+n_{k_3}) \right] \right\} \nonumber\\
&&                                                \label{retaredsigma}
\end{eqnarray}
The linearized equation of motion (\ref{finaleqofmot}) can now be
solved by the Laplace transform. With the boundary conditions
$\phi_{\vec p} (t=0) = \phi_{i; \vec p} \, ; \dot{\phi}_{\vec p}(t=0)
= 0$ and denoting the Laplace transforms of $\phi_{\vec p}(t)\,,
\Sigma_{\vec p}(t)$ by $\phi_{\vec p}(s)\,, \Sigma_{\vec p}(s)$
respectively (here $s$ is the Laplace transform variable) we find
\begin{equation}
\phi_{\vec p}(s)=\frac{\phi_{i;\vec p}\; s}{s^2+\omega^2_{p}
+\Sigma_{\vec p}(s)}                                     \label{lapla}
\end{equation}
The time evolution is found by inverse Laplace transform, i.e. by
integration in the complex s-plane along the Bromwich contour
\begin{equation}
\phi_{\vec p}(t)= \int^{i\infty+\epsilon}_{-i\infty+\epsilon}
e^{st}\phi_{\vec p}(s) \frac{ds}{2\pi i}              \label{bromwich}
\end{equation}
\noindent with $\epsilon$ to the right of the real part of all the
singularities of the Laplace transform $\phi_{\vec p}(s)$. The Laplace
transform of the retarded self-energy has the following spectral
representation
\begin{equation}
\Sigma_{\vec{p}} (s)=  -\int \frac{2\, p_0\,
\rho(p_0,{\vec{p}}; T)}{s^2+p_0^2}
dp_o \;.                                            \label{dispersion}
\end{equation}
where $\rho(p_0,{\vec{p}}; T) $ is the spectral density, given to two-
loop order by
\begin{eqnarray}
\rho(\vec{k},p_0;T)&=& {\lambda^2 \over 6} \int {d^3 k_1 \over
(2 \pi)^3} \; {d^3 k_2 \over (2 \pi)^3} \; {d^3 k_3 \over (2 \pi)^3}
\; {1 \over 8 \omega_1 \omega_2 \omega_3 } \; \left[ (2 \pi)^3
\delta^3 ({\vec{k}}-{\vec{k}}_1 - {\vec{k}}_2- {\vec{k}}_3) \right]
\nonumber \\
&& \left\{ \delta (p_0-\omega_{k1}-\omega_{k2}-\omega_{k3}) \left(
(1+n_{k1})\, (1+n_{k2}) \, (1+n_{k3}) -n_{k1} \, n_{k2} \, n_{k3}
\right) \right. \nonumber \\
&&\left. + 3 \, \delta (p_0+ \omega_{k1}-\omega_{k2}-
\omega_{k3}) \left( n_{k1} \, (1+n_{k2}) \,(1+ n_{k3}) -(1+n_{k1}) \,
n_{k2} \, n_{k3} \right) \right\} \nonumber\\
                                                      \label{spectral}
\end{eqnarray}
This spectral density displays two different type of contributions
\cite{weldon,blaizot,wang}: the first one (corresponding to the first
$\delta$-function) corresponds to the production process $\phi
\rightarrow \phi \phi \phi$ with the Boltzmann weight
$(1+n_{k1})(1+n_{k2})(1+n_{k3})$ minus the recombination $\phi \phi
\phi \rightarrow \phi$ in the medium with the weight $n_{k1} n_{k2}
n_{k3}$. The second contribution (second $\delta$-function)
corresponds to the scattering process $\phi \phi \rightarrow \phi
\phi$ with weight $n_{k1}(1+n_{k2})(1+n_{k3})$ minus the inverse
process with weight $n_{k1} n_{k2} (1+n_{k3})$. The self-energy
$\Sigma_{\vec p}(s)$ has an imaginary part along the imaginary axis in
the s-plane which can be read off from the spectral representation:
\begin{equation}
\Sigma_{I,{{\vec{p}}}} (i\omega \pm 0^+) = \pm \pi \; {\rm{sign}} \,
(\omega) \left[ \rho( |\omega|, {{\vec{p}}},T)- \rho (
-|\omega|,{{\vec{p}}},T ) \right]                    \label{impartdef}
\end{equation}
where the ${\rm{sign}}(\omega)$ reflects the retarded nature
\cite{kapusta,wang}. To two-loop order we find the imaginary part to
be given by:
\begin{eqnarray}
\Sigma_{I, {\vec{k}}} (i\omega+0^+)&={\lambda^2 \over 6} \, \pi
{\rm{sign}} (\omega)&\int {d^3 k_1 \over (2 \pi)^3} \, {d^3 k_2 \over
(2 \pi)^3} \, {d^3 k_3 \over (2 \pi)^3} \, {1 \over 8 \omega_1
\omega_2 \omega_3 } \; \left[ (2 \pi)^3 \delta^3 ({{\vec{k}}} -
{\vec{k}}_1 - {\vec{k}}_2- {\vec{k}}_3) \right]           \nonumber \\
&& \qquad\qquad \times
\big[ \delta( |\omega| -\omega_{k1} -\omega_{k2}
-\omega_{k3}){\cal N}_1                                   \nonumber \\
&& \qquad\qquad\qquad
+ 3 \delta( |\omega| +\omega_{k1} -\omega_{k2}
-\omega_{k3}){\cal N}_2                                   \nonumber \\
&& \qquad\qquad\qquad\qquad
+ \delta( -|\omega| -\omega_{k1} -\omega_{k2}
-\omega_{k3}) {\cal N}_3                                  \nonumber \\
&& \qquad\qquad\qquad\qquad\qquad
+ \delta(- |\omega| +\omega_{k1} -
\omega_{k2} -\omega_{k3}) {\cal N}_4 \big]
                                                     \label{imaginary}
\end{eqnarray}
where
\begin{eqnarray}
{\cal N}_1 &=& (1+n_{k1}) \, (1+n_{k2})\, (1+n_{k3}) -n_{k1} \, n_{k2}
\, n_{k3}                                                 \nonumber \\
{\cal N}_2 &=& n_{k1} \, (1+n_{k2}) \, (1+n_{k3}) -(1+n_{k1})\, n_{k2}
\, n_{k3}                                                 \nonumber \\
{\cal N}_3 &=& n_{k1} \, n_{k2} \, n_{k3} -(1+n_{k1}) \, (1+n_{k2}) \,
(1+n_{k3})                                                \nonumber \\
{\cal N}_4 &=& (1+ n_{k1}) \, n_{k2} \, n_{k3} -n_{k1} \, (1+n_{k2})
\, (1+n_{k3})                                                \label{ocupanumb1}
\end{eqnarray}
This result coincides with that found by Wang and Heinz\cite{wang}.
The first term gives a contribution above the three particle threshold
at $\omega > 3m_R(T)$, the third term vanishes identically, the second
and fourth term give the collisional contribution and have support all
along the imaginary axis\cite{wang}. Therefore along the imaginary
axis we find

\begin{equation}
\Sigma_{\vec{k}}(i\omega\pm 0^+)=\Sigma_{R,{\vec{k}}}(\omega)\pm
i\Sigma_{I,{\vec{k}}}(\omega)                          \label{sigphys}
\end{equation}

The real part can be obtained by a dispersion relation. It is very
difficult to obtain in general, but we find that it is quadratically
and logarithimically divergent. These divergences can be subtracted
away from the dispersion integral and the subtraction absorbed in a
mass and wave function renormalization.

In order to invert the Laplace transform and obtain the real-time
evolution, we need to understand the analytic structure of $\phi_{\vec
p}(s)$ in the complex s-plane. Isolated single particle poles would be
at the values $s=\pm i\Omega(\vec p,T)$ with $\Omega(\vec p,T)$ the
solutions of
\begin{equation}
-\Omega^2(\vec p,T)+|\vec p |^2+M^2_{\sigma}(T)+\Sigma(s=\pm i \Omega,
\vec p, T)=0                                     \label{polecondition}
\end{equation}
Since the self-energy is complex along the imaginary axis, such a pole
solution would be complex. However it is easy to see that there is no
solution to (\ref{polecondition}) in the first (physical) Riemann
sheet. The pole moves off into the second (unphysical) Riemann sheet
as corresponds to a ``resonance'' (or quasiparticle pole)\cite{B3}.
The discontinuity of the Laplace transform across the imaginary axis
is given by:
\begin{equation}
\phi(\vec p, s=i\omega+0^+)-\phi(\vec p, s=i\omega-0^+)=
\frac{\phi_{i;\vec p}\; \; 2i\omega\,\Sigma_I(\omega,\vec p,T)}
{[\omega^2 -|\vec p|^2-M_{\sigma}^2- \Sigma_R(\omega,\vec p,T)]^2+
\Sigma_I(\omega,\vec p,T)^2}.                    \label{discontinuity}
\end{equation}
This discontinuity is the only singularity in the (first Riemann
sheet) complex s-plane and it vanishes at the origin. This fact allows
us to deform the integration countour to wrap around the cut.

Finally for the time evolution we obtain
\begin{equation}
\phi_{\vec p}(t)=\frac{2 \phi_{i;\vec p}}{\pi} \,  \int_{0}^\infty
{{\omega \Sigma_I(\omega,\vec p;T) \cos(\omega t)\,d\omega}\over
{[\omega^2-\omega^2_p- \Sigma_R(\omega,\vec p;T)]^2+
\Sigma_I(\omega,\vec p; T)^2}}                        \label{unstable}
\end{equation}
with $\omega^2_p$ given by (\ref{sigmafreq}).

Setting $t=0$ in the above expression leads to the sum rule
\begin{equation}
\frac{2}{\pi} \,  \int_{0}^\infty
\frac{\omega\Sigma_I(\omega,\vec p;T) \, d\omega}
{[\omega^2-\omega^2_p- \Sigma_R(\omega,\vec p;T)]^2+
\Sigma^2_I(\omega,\vec p; T)} =1                       \label{sumrule}
\end{equation}
which has been previously obtained by Pisarski\cite{pisasumrule} in a
rather different manner.

For very weak coupling the integrand in (\ref{unstable}) features a
narrow resonance at $\omega^2 = \omega^2_p + {\cal O}(\lambda^2)$, and
can be approximated by a Breit-Wigner form leading to the real-time
evolution\cite{B3}
\begin{equation}
\phi_{\vec p}(t)= \phi_{i; \vec p} Z[\vec p;T] e^{-\Gamma(\vec p; T)
t} \cos(\omega_pt+\alpha)                             \label{realtime}
\end{equation}
with
\begin{eqnarray}
\Gamma(\vec p;T) &\approx & Z[\vec p,T]
\frac{\Sigma_I(\omega=\omega_p, \vec p; T)}{2\omega_p}
                                                  \label{gammarate} \\
Z[\vec p;T] &=& \left[\left.1- \frac{\partial \Sigma_R(\omega,
\vec p, T)} {\partial \omega^2}\right|_{\omega=\omega_p}\right]^{-1}=
1+ {\cal O} (\lambda^2)                      \label{wavefunctionren}\\
\alpha &=& - \left. \frac{\partial \Sigma_I(\omega, \vec p, T)}
{\partial \omega^2}\right|_{\omega=\omega_p}              \label{alfa}
\end{eqnarray}
where $Z[\vec p;T]$ is the wave function renormalization defined on
shell. To the order that we are working, this will be set to one in
(\ref{gammarate}). $\Gamma(\vec p;T)$ is the {\em collisional}
relaxation rate. The Breit-Wigner approximation, however, is valid
only for times $t \leq \frac{1}{\Gamma} \ln\left(\frac{\omega_p}
{\Gamma}\right)$. For longer times the fall-off is with a power law
determined by the behavior of the spectral density at small frequency
\cite{B3}.

The calculation of the decay rate $\Gamma(\vec p;T)$ is in general
very difficult for arbitrary momentum and temperature (see
however\cite{wang}). For $p \ll M_R(T) \ll T$ we can approximate it by the
zero momentum limit, which can be calculated relatively easily. The
only contribution to the imaginary part on-shell is given by the
second and fourth terms (collisional terms) in (\ref{imaginary}).

We find for $T\gg \sqrt{\lambda}T \gg |m_R|$:
\begin{eqnarray}
\Gamma(\vec p=0; T) = \frac{\lambda^{3/2}T}{64\sqrt{24\pi}}
                                                  \label{zeromomrate}
\end{eqnarray}
The same result had been obtained previously by many authors
\cite{parwani,elmfors,wang}

\subsection{\bf $T<<T_c$: Broken symmetry case}

In this section we consider the linear relaxation of fluctuations
around the broken symmetry state, for non-zero temperature but below
the critical value. In this case the scalar field acquires an
expectation value $v$ which at tree level is given by
\begin{equation}
v = \sqrt{\frac{6|m^2_0|}{\lambda}}                        \label{vev}
\end{equation}
The small amplitude fluctuations around the broken symmetry state have
a tree level mass squared given by
\begin{equation}
M^2_0  =m^2_0-\frac{\lambda v^2}{2}=  {2 |m^2_0|}     \label{treemass}
\end{equation}

We now write the field $\Phi^{\pm}$ as
\begin{equation}
\Phi^{\pm}(\vec x, t)= v+\phi(\vec x,t)+\psi^{\pm}(\vec x, t)\; .
                                                  \label{broksymsplit}
\end{equation}
We will {\em not} fix $v$ at the tree level value given by eq.
\ref{vev}, but instead the v.e.v will be determined order by order in
the perturbative expansion by requiring that the linearized equation
of motion for $\phi(\vec x,t)$ be homogeneous. Furthermore as in the
previous case we will write $M^2_0= M^2_R(T) + \delta m^2(T)$ and
treat $\delta m^2(T)$ as a counterterm. The microscopic time scale is
determined by $M^{-1}_R(T)$. The condition $\langle \psi^{\pm}(\vec
x,t)\rangle=0$ implemented order by order leads to the equation of
motion for $\phi(\vec x, t)$ as before\cite{B3}. In the broken
symmetry phase there are new Feynman diagrams that contribute an
absorptive part to the self-energy at {\em one-loop} order which are
proportional to the v.e.v. Now there are new vertices in the
``shifted'' Lagrangian density. Those that contribute to the
linearized equation of motion are obtained from the part of the
Lagrangian density given by
\begin{equation}
{\cal L}_v = \left\{ v\psi^+ (m^2_0 - \frac{\lambda}{6}v^2)
-\frac{\lambda v}{6}(3 \phi (\psi^+)^2 + (\psi^+)^3)\right\} - \left\{
\psi^+ \rightarrow \psi^-\right\}                      \label{lagranv}
\end{equation}

To one-loop order there are two tadpole terms that contribute to the
equation of motion. One is independent of $\phi$ and is absorbed in a
renormalization of $v$ (since it gives an inhomogeneous contribution
to the equation of motion). The other is proportional to $\phi$ and is
cancelled by the mass counterterm leading to the ``gap equation''
(\ref{gapeq}, \ref{gapeqren}), but in the present case $M_R(T) > T \gg
\sqrt{\lambda} T$. The remaining one-loop diagrams contributing to the
self-energy with an absorptive part are shown at the top of figure 2.
We obtain the equation of motion given by eq. (\ref{eqnofmot}) with a
new one-loop contribution to the retarded self-energy as well as the
two-loop contribution obtained in the previous section and given by
eq. (\ref{sigmaret}). We find the one loop contribution to be given by
(see also\cite{B3,lastpap})
\begin{equation}
\Sigma^{v}_{ret} (\vec{x}-\vec{x'},t-t') = -i \frac{\lambda^2 v^2}{2}
[(-iG^>(\vec{x}-\vec{x'},t-t'))^2 - (-iG^< (\vec{x}-\vec{x'},t-t'))^2]
\Theta(t-t')                                      \label{sigmaret1lup}
\end{equation}
with the following spatial Fourier transform:
\begin{eqnarray}
\Sigma^{v}_{\vec p}(t-t')&=&-\frac{\lambda^2 v^2}{2}
\int\frac{d^3k}{(2\pi)^3} \frac{1}{2\omega_{k} \omega_{\vec k+ \vec
p}}\{ (1+2 n_{k})\sin[(\omega_{\vec k+\vec p}+\omega_{k})(t-t')]
\nonumber \\
&&\quad+2 n_{k} \sin[(\omega_{\vec k+\vec p}-\omega_{k})(t-t')]\}
                                                  \label{sigma1lupk}
\end{eqnarray}

The Laplace transform can again be written in a spectral
representation as given by (\ref{dispersion}), and we find the
one-loop contribution to the spectral density to be
\begin{equation}
\rho^{v}(\vec p, p_o) = \frac{\lambda^2 v^2}{4}\int \frac{d^3k}{(2
\pi)^3} \frac{1}{2 \omega_{\vec k}\omega_{\vec k+ \vec p}} \{
(1+2n_{k}) \delta(p_o-\omega_{\vec k}-\omega_{\vec k + \vec p})-
2n_{k} \delta(p_o - \omega_{\vec k}+ \omega_{\vec k+\vec p} ) \}
                                                      \label{specdens}
\end{equation}
The imaginary part is obtained from (\ref{impartdef}), and after
analyzing the kinematical regions we find for $\omega >0$ (for $\omega
<0$ there is a sign change) (see also\cite{lastpap})
\begin{eqnarray}
Im\Sigma^{v}_{\vec p}( i\omega + 0^+,T) & = &
\left\{\frac{\lambda^2 v^2}{32\pi}\sqrt{1-\frac{4M^2_R(T)}
{\omega^2-p^2}} + \frac{\lambda^2 v^2 T}{16\pi p}
\ln\left[\frac{1-e^{-\beta \omega_p^+}} {1-e^{-\beta \omega_p^-}}
\right] \right\} \Theta(\omega^2-p^2-4M^2_R(T)) \nonumber \\
&&+ \frac{\lambda^2 v^2 T}{16\pi p} \ln\left[\frac{1-e^{-\beta
\omega_p^+}} {1-e^{-\beta \omega_p^-}}\right] \Theta(p^2-\omega^2)
                                                   \label{impar1lup}\\
\omega_p^{\pm}&=&\left|\frac{\omega}{2}\pm \frac{p}{2}
\sqrt{1-\frac{4M^2_R(T)}{\omega^2-p^2}}\right|\, .\label{omegaplusmin}
\end{eqnarray}
The processes that contribute to the first term above are the creation
of two mesons in the medium with Boltzmann weight $(1+n_{\vec k})
(1+n_{\vec k + \vec p})$ minus the inverse process with weight
$n_{\vec k}n_{\vec k + \vec p}$, whereas the processes that contribute
to the second term are scattering off quanta in the bath with weight
$(1+n_{\vec k}) n_{\vec k + \vec p}$ minus its inverse with weight
$(1+n_{\vec k + \vec p})n_{\vec k}$ \cite{weldon,blaizot}. Notice that
whereas the first term displays the usual two particle threshold, the
second term is reminiscent of Landau damping in that it only has
support below the light cone, for $|p| > |\omega| $
\cite{iancu,blaizot}. Neither term contributes on-shell. However there
is a new contribution of ${\cal O}(\lambda)$ to mass renormalization
and a finite contribution of the same order to wave function
renormalization. In terms of the renormalized mass, the position of
the resonance obtains a correction of ${\cal O}(\lambda)$ but to this
order the width remains the same since the imaginary part on shell is
already of ${\cal O}(\lambda ^2)$.

The results of this section also apply if there is a {\em trilinear}
coupling of the scalar field $\Phi$ to lighter scalars of the form $g
\Phi \chi^2$. In this case the result (\ref{impar1lup}) applies (with
$\lambda^2 v^2 \rightarrow 4 g^2$) but with the mass $M_R(T)$ in the
argument of the square roots in equations (\ref{impar1lup},
\ref{omegaplusmin}) replaced by $m_{\chi}(T)$ the (thermal) mass of
the $\chi$ particle in the loop. If $M_R(T) > 2 m_{\chi}(T)$, decay is
kinematically allowed (on-shell) and the scalar particle will acquire
a contribution from the imaginary part of the self-energy on shell
from the decay process $\Phi \rightarrow \chi \chi$ minus the inverse
(recombination) process $\chi \chi \rightarrow \Phi$. The decay (minus
recombination) width for weak coupling is therefore(see also\cite{lastpap})
\begin{eqnarray}
\Gamma_{\Phi \leftrightarrow \chi\chi}&\approx& \frac{g^2}{16\pi
\omega_p}\sqrt{1-\frac{4m^2_{\chi}(T)}{M^2_R(T)}} + \frac{g^2 T}{8\pi
p \omega_p} \ln\left[\frac{1-e^{-\beta W_p^+}} {1-e^{-\beta
W_p^-}}\right]                                    \label{decayrate} \\
W_p^{\pm}&=& \left|\frac{\omega_p}{2}\pm \frac{p}{2}
\sqrt{1-\frac{4m^2_{\chi}(T)}{M^2_R(T)}}\right|\ .    \label{Wplusmin}
\end{eqnarray}

Now the collisional and the decay width add up. The Landau damping
term does not contribute to the width of the particle {\em directly},
because it has support only below the light cone. Notice, however,
that the Landau damping contribution to (\ref{decayrate}) is the same
as the finite temperature contribution above the two particle
threshold in the first term of (\ref{impar1lup}). The sum rule
(\ref{sumrule}) ensures that the Landau damping contribution is
``borrowed'' from the spectral density above the two particle
threshold and thus contributes {\em indirectly} to the width of the
particle and to the wave function renormalization (the weight of the
quasiparticle pole).

The main purpose of studying the broken symmetry phase is to compare
to the kinetic description and recognize the terms corresponding to
decay and recombination processes in the evolution equation for the
distribution function to be studied in the next section.

\section{Kinetic theory}

In this section we obtain the evolution equations for the distribution
functions of quasiparticles, including off-shell effects. Here we
present an alternative to the usual derivation in which correlation
functions are written in terms of relative and ``center of mass''
space-time coordinates and the Wigner transform in the relative
coordinates is performed. In this approach the Boltzmann equation is
obtained in a gradient expansion assuming that the dependence on
``center of mass'' coordinates is weak. We will not assume such a
situation, but instead analyze which are the relevant time scales over
which a coarse graining procedure must be implemented.

Let us begin by writing the Hamiltonian of the theory as
\begin{eqnarray}
H & = & H_0 + H_{int}                                \label{hsplit} \\
H_0 & = & \frac{1}{2} \int d^3x \left\{\Pi^2 +(\nabla \Phi)^2 +
M^2_R(T) \Phi^2\right\} \ ,                              \label{hfree}
\end{eqnarray}
Where the mass counterterm has been absorbed in the interaction. The
interaction part $H_{int}$ will depend on whether $T>>T_c$ or $T<T_c$
is considered. The part of the Hamiltonian $H_0$ describes free
quasiparticles of renormalized finite temperature mass $M_R(T)$ and is
diagonalized in terms of creation and annihilation of free
quasiparticle operators $a^{\dagger}_k \; ; a_k$ respectively. The
pole mass of the quasiparticles will acquire corrections in
perturbation theory, but these will remain perturbatively small at
large temperatures.

With this definition, the lifetime and weight of the quasiparticles
will be a consequence of interactions. In this manner, the hard
thermal loops which in this theory amount to local terms, have all
been absorbed in the definition of the thermal mass, which guarantees
that the microscopic time scales are explicit in the quasiparticle
hamiltonian. Such an approach has been previously advocated by
Lawrie\cite{lawrie}.

Now consider the case in which at an initial time (to be taken as
$t=t_0$) the density matrix is diagonal in the occupation number of
the free quasiparticles, but out of equilibrium, with initial
occupation numbers $N_k(t_0)$. A comment is in order here. In an
initial situation out of equilibrium the temperature is not a
meaningful quantity and conceptually $M_R(T)$ in the above Hamiltonian
$H_0$ is not well defined. However we are interested in small
departures from equilibrium as is implicit in a kinetic description,
thus the value of the temperature in the resummed mass term should be
thought of as the final equilibrium temperature. In particular for the
relaxation time approximation a typical situation considered is that
only few modes are out of equilibrium whereas all other are in
equilibrium and serve as the ``bath''. In this case the temperature T
in the mass parameter is that of the modes in equilibrium. This will
become clear in the relaxation time approximation discussed later.

The Heisenberg field operators at time $t$ are now written as
\begin{eqnarray}
\Phi ( \vec{x},t)&=&\frac{1}{\sqrt{\Omega}} \sum_k \; \Phi_k(t) \;
e^{i \vec{k} \cdot \vec{x}}\;, \quad \quad \Phi_k(t)= \frac{1}{\sqrt{2
\omega_k }} \left( {a}_k (t) + {a}^{\dagger}_{-k} (t) \right)
                                                          \nonumber \\
\Pi ( \vec{x},t)&=& \frac{1}{\sqrt{\Omega}} \sum_k \; \Pi_k(t) \; e^{i
\vec{k} \cdot \vec{x}} \;, \quad \quad \Pi_k(t) = {-i \omega_k \over
\sqrt{2 \omega_k}} \left( {a}_k (t) - {a}^{\dagger}_{-k} (t) \right)
                                              \label{initialexpansion}
\end{eqnarray}
where $\Omega$ is the spatial volume and the time evolution of the
creation and annihilation operators is through the time evolution
operator. $\omega^2_k $ are the same as in (\ref{bosefactor}).

The expectation value of particle number operator $N_k$ can be
expressed in terms of the fields $\Phi_k $ and that of the conjugate
momentum $\Pi_k $ as follows,
\begin{equation}
N_k (t) = \langle {a}^{\dagger}_{k} (t) \;{a}_k (t) \rangle = {1 \over
2\; \omega_k }\langle \Pi_{k}(t) \; \Pi_{-k} (t) \; + \omega^2_k \;
\Phi_k (t) \; \Phi_{-k} (t) \; \rangle -{1 \over 2}  \label{ocupation}
\end{equation}
where the bracket $\langle \cdots \rangle$ means an average over the
gaussian density matrix defined by the initial occupation numbers
$N_k(t_0)$. The time-dependent distribution (\ref{ocupation}) is
interpreted as the quasiparticle distribution function.

As in our study of relaxation in the previous section, we examine
separately the cases $T>>T_c$ and $T<< T_c$ since the interaction
vertices are different (we could study the situation in the most
general case, but prefer to study in detail the cases separately to
illustrate clearly the different processes).

\subsection{\bf $T>> T_c$}

The interaction Hamiltonian in this case is given by
\begin{equation}
H_{int} =\frac{\lambda}{4!} \frac{1}{\Omega}\sum_{\vec k1,\vec k2,
\vec k3,\vec k4} \Phi_{\vec k1}\Phi_{\vec k2}\Phi_{\vec k3}\Phi_{\vec
k4}\delta_{\vec k1 + \vec k2+ \vec k3 + \vec k4}+ \delta m^2(T)
\sum_{\vec k}\Phi_{\vec k} \Phi_{-\vec k}            \label{intTlarge}
\end{equation}
Taking the derivative of $N_k (t)$ with respect to time and using the
Heisenberg field equations, we find
\begin{eqnarray}
{\dot{N}}_k (t) ={1 \over 2 \, \omega_{k}} \Bigg[ -{\lambda \over 6} \,
\langle (\Phi)^3_{k} (t) \, \Pi_{-k} (t)&+&\Pi_{k} (t) \, (\Phi
)^3_{-k} (t) \, \rangle  \nonumber \\
&-& \delta m^2(T) \, \langle \Phi_{k}(t) \, \Pi_{-k}(t) +\Pi_{k}(t) \,
\Phi_{-k}(t) \rangle \Bigg]                              \label{nkdot}
\end{eqnarray}
where we use the compact notation:
\begin{equation}
(\Phi)^3_{k} (t) \equiv {1 \over \Omega} \sum_{k_1, k_2, k_3}
\delta_{k_1+k_2+k_3-k,0} \; \Phi_{k_1} (t) \; \Phi_{k_2} (t) \;
\Phi_{k_2} (t)
\end{equation}

In a perturbative expansion care is needed to handle the canonical
momentum ($\Pi = \dot{\Phi}$) and the scalar field at the same time
because of Schwinger terms. This ambiguity is avoided by defining
\begin{equation}
\langle \Pi_{k} (t) \; (\Phi)^3_{-k} (t) \rangle= tr \left[
\rho (t_0) \; \Pi_{k} (t)\;(\Phi)^3_{-k}(t) \right]
\equiv \lim_{ t \rightarrow t'} {\partial \over \partial\; t'}
tr \left[(\Phi^+)^3_{-k} (t) \; \rho (t_0) \; \Phi^-_{k} (t') \right]
\end{equation}
Where we used the cyclic property of the trace and the $\pm$
superscripts for the fields refer to field insertions obtained as
variational derivatives with respect to sources in the forward time
branch ($+$) and backward time branch ($-$) in the non-equilibrium
generating functional\cite{B3}.

We now use the canonical commutation relation between $\Pi$ and $\Phi$
and define the mass counterterm $\delta m^2(T) = (\lambda /6)
\Delta(T)$ to write the above expression as
\begin{eqnarray}
{\dot{N}}_k (t) & = & -\frac{\lambda}{12\omega_k}\left\{2
\frac{\partial}{\partial t'} \left[ \langle
(\Phi^+(t))^3_{k}\Phi^-_{-k}(t')\rangle + \Delta(T)\langle \Phi^+_k(t)
\Phi^-_{-k}(t')\rangle \right]_{t = t'}\right. \nonumber \\
& + &\left. {3i}\left[\frac{1}{\Omega} \sum_k \langle \Phi^+_k(t)
\Phi^-_{-k}(t)+ \frac{ \Delta(T)}{3} \right] \right\}   \label{nkdot1}
\end{eqnarray}

The right hand side of eq. (\ref{nkdot1}) can be obtained in weak
coupling expansion in $\lambda$. Such a perturbative expansion is in
terms of the non-equilibrium Green's functions (\ref{gplusplus}-
\ref{gsmaller}) with the basic Green's functions given by
(\ref{ggreat}, \ref{gsmall}) but with the non-equilibrium occupation
number $N_k(t_0)$ replacing the equilibrium one $n_k$. An important
point to notice is that these Green's functions include the proper
microscopic scales as the contribution of the hard thermal loops have
been incorporated by summing the tadpole diagrams. The propagators
entering in the calculations are the resummed propagators. The terms
with $\Delta(T)$ are required to cancel the tadpoles to all orders.

As will be discussed in detail below, such an expansion will be
meaningful for times $t \ll \tau_r =| ( N_k (t)/ {\dot{N}}_k (t))| $,
where $\tau_r$ is the relaxation time scale for the non-equilibrium
distribution function. For small enough coupling we expect that
$\tau_r$ will be large enough so that there is a wide separation
between the microscopic and the relaxation time scales that will
warrant such an approximation (see discussion below). At order ${\cal
O}(\lambda)$ the right hand side of (\ref{nkdot1}) vanishes
identically. This is a consequence of the fact that the initial
density matrix is diagonal in the occupation number. Out of
equilibrium, when the {\em equal time} correlation functions in
(\ref{nkdot1}) are time dependent, the mass counterterm will acquire a
(weak) time dependence through the non-equilibrium distribution
functions, thus $\Delta(T,t)$ will be a slowly varying function of
time on a scale determined by $\tau_r$. Thus from the formidable
expression (\ref{nkdot1}) only the first term remains after the hard
thermal loop resummation and we find one of our main results:
\begin{equation}
{\dot{N}}_k (t) = -\frac{\lambda}{6\omega_k} \frac{\partial} {\partial
t'} \left[ \langle (\Phi^+(t))^3_{k}\Phi^-_{-k}(t')\rangle \right]_{t
= t'}                                                 \label{ndotequa}
\end{equation}
with the understanding that no tadpole diagrams contribute to the
above equations as they are automatically cancelled by the terms
containing $\Delta(T)$ in (\ref{nkdot1}).

To two-loop order, the diagrams that contribute to (\ref{ndotequa})
are shown in figure 1. We find the following expression for the
evolution of the distribution function:
\begin{eqnarray}
{\dot{N}}_k (t) &=& {\lambda^2 \over 3} \, {1 \over 2 \, \omega_k} \,
\int {d^3 k_1 \over (2 \pi)^3} \, {d^3 k_2 \over (2 \pi)^3} \, {d^3
k_3 \over (2 \pi)^3} \, {1 \over 8 \omega_{k1} \omega_{k2} \omega_{k3}
} \, (2 \pi)^3 \delta^3 ({\vec{k}}-{\vec{k}}_1 - {\vec{k}}_2-
{\vec{k}}_3)                                              \nonumber \\
&&\qquad \qquad \times \Bigg\{
\left[\frac{ \sin \left[ ( \omega_k  + \omega_{k1}  + \omega_{k2}+
\omega_{k3} ) (t-t_0) \right]}{ \omega_k + \omega_{k1} + \omega_{k2}+
\omega_{k3} } \right]{\cal N}_1(t_0)                      \nonumber \\
&&\qquad \qquad \qquad \;
+ \left[\frac{ 3\,\sin \, \left[ ( \omega_k + \omega_{k1} +
\omega_{k2}- \omega_{k3} )(t-t_0) \right]}{ \omega_k + \omega_{k1} +
\omega_{k2}- \omega_{k3} } \right]{\cal N}_2(t_0)         \nonumber \\
&&\qquad \qquad \qquad \qquad \;
+\left[\frac{3\,\sin \, \left[ ( \omega_k + \omega_{k1}
-\omega_{k2}- \omega_{k3} )(t-t_0) \right]}{\omega_k + \omega_{k1}
-\omega_{k2}- \omega_{k3} }\right]{\cal N}_3(t_0)         \nonumber \\
&&\qquad \qquad \qquad \qquad \qquad \;
+ \left[\frac{ \sin \, \left[ ( \omega_k -\omega_{k1}
-\omega_{k2}- \omega_{k3} ) (t-t_0) \right]}{ \omega_k -\omega_{k1}
-\omega_{k2}- \omega_{k3} } \right]{\cal N}_4(t_0) \Bigg\}
                                                         \label{boltz}
\end{eqnarray}
where
\begin{eqnarray}
{\cal N}_1(t) &=& (1+ N_k(t))\,(1+ N_{k1}(t))\, (1+ N_{k2} (t))\,
(1+ N_{k3} (t))- N_k (t) \, N_{k1} (t) \, N_{k2} (t) \;
N_{k3}(t)                                                 \nonumber \\
{\cal N}_2(t) &=& (1+ N_k(t) ) \,(1+ N_{k1} (t)) \, (1+ N_{k2}
(t)) \, N_{k3} (t)- N_k (t) \, N_{k1} (t) \, N_{k2} (t) \,
(1+N_{k3}(t) )                                            \nonumber \\
{\cal N}_3(t) &=& (1+ N_k (t) )\,(1+ N_{k1} (t)) \, N_{k2} (t) \,
N_{k3} (t)- N_k (t) \, N_{k1} (t) \, (1+ N_{k2} (t)) \,
(1+N_{k3}(t) )                                            \nonumber \\
{\cal N}_4(t) &=& (1+ N_k (t) )\, N_{k1} (t) \, N_{k2} (t) \,
N_{k3} (t)- N_k (t) \, (1+ N_{k1} (t)) \, (1+ N_{k2} (t)) \,
(1+N_{k3}(t) )                                             \label{ocupanumber22}
\end{eqnarray}

There are several noteworthy features of this expression:

{\bf i)} $\dot{N}_k(t)$ vanishes at $t=t_0$. This is a consequence of
the choice of the initial density matrix which is diagonal in the
$N_k(t_0)$ basis, so that the occupation number operator commutes with
the density matrix at $t=t_0$.

{\bf ii:)}
consider times much larger than the time scales implicit in the
arguments of the sine functions, typically $t-t_0 \gg M^{-1}_R(T)$.
Assuming that the rate of change of the occupation numbers is very
slow so that during this time the occupation numbers have not changed
significantly, namely, $\tau_r \approx N_k(t_0)/ \dot{N}(t_0) \gg
(t-t_0) \gg M^{-1}_R(T)$, we can approximate
\begin{equation}
\frac{\sin(W(t-t_0))}{W} \approx \pi \delta(W)      \label{goldenrule}
\end{equation}
where $W$ is any of the combinations entering in the arguments of the
sine functions above. This approximation is the same as that invoked
in time-dependent perturbation theory leading to Fermi's Golden Rule.
Under the assumptions leading to the approximation (\ref{goldenrule})
we recognize that each term in the expression (\ref{boltz}) can be
identified with a term having the same structure in the imaginary part
of the self energy given by eq. (\ref{imaginary}), {\em including
off-shell processes} when the non-equilibrium occupation numbers
$N_k(t_0)$ are replaced by $n_k$. The first term describes the
creation of four particles minus the destruction of four particles in
the plasma, the second and fourth terms describe the creation of three
particles and destruction of one minus destruction of three and
creation of one, the third term is the {\em scattering} of two
particles off two particles and is the usual Boltzmann term. If the
$N_k(t_0)$ coincide with the equilibrium distribution functions $n_k$
then {\em all the terms} in eq. (\ref{boltz}) vanish identically and
${\dot N}_k=0$, including the off-shell terms as a consequence of the
energy conservation delta functions.

{\bf iii)} Using the approximation (\ref{goldenrule}) and assuming
that all the modes but the one with wavector $k$ are in equilibrium,
and the mode with wavevector $k$ is slightly off-equilibrium, that is
$N_{ki}= n_{ki}\, ; N_k= n_k+\delta n_k(t_0)$ to linear order in
$\delta n_k(t_0)$, we obtain to linear order in the $\delta n_k(t_0)$
the ``relaxation time approximation''
\begin{equation}
\frac{d \delta n_k}{dt} = -\frac{\Sigma_I(\omega_k)}{\omega_k}
\delta n_k(t_0)                                    \label{linearboltz}
\end{equation}
where $\Sigma_I(\omega_k)$ is given by eq. (\ref{imaginary}). Thus the
relation between the rate of change of the non-equilibrium
distribution function in the relaxation time approximation and the
``damping rate'' of quasiparticles becomes explicit.

{\bf iv:)} Only the third (scattering) term in the evolution equation
(\ref{boltz}) conserves the total number of particles $N = \sum_k
N_k$. The other three terms are a consequence of the fact that the
interaction {\em does not} conserve the particle number and this gives
rise to off-shell processes that change the particle number. However
as explained above, these processes only contribute on time scales
comparable to the microscopic scale. These particle-number-changing
processes produce departures from chemical equilibrium on short time
scales.

Obviously, equations (\ref{boltz}, \ref{linearboltz}) describe only
the early time evolution of the distribution functions. They neglect
the change in the initial occupation numbers and, as they stand,
cannot be extended to long times. Consider however the following
procedure: assume that there is a wide separation of time scales in
the sense that $\tau_r \gg M^{-1}_R(T)$, and consider integrating
(\ref{boltz}) (or the linearized version (\ref{linearboltz})) in a
time interval from $t=t_0$ to $t_0+\Delta t$ such that $M^{-1}_R(T)
\ll \Delta t \ll \tau_r$. In this time interval we can approximate the
sine functions by energy conserving delta functions using
(\ref{goldenrule}) keeping the occupation numbers at their initial
value at $t=t_0$. At the end of this interval ``reset'' all the
occupation numbers to the values $N_k(t_0+\Delta t)$, and furthermore
reduce the density matrix at this time ($t_0+\Delta t$) to the
diagonal elements in the basis of the occupation numbers at this time.
Then the Green's functions will look exactly the same as those given
at the initial time $t=t_0$ but in terms of the occupation numbers at
$t=t_0+\Delta t$. In this ``coarse grained'' density matrix there are
no off-diagonal matrix elements of the creation and destruction
quasiparticle operators at $t=t_0+\Delta t$. In particular, this
reduction of the density matrix neglects new correlations of the form
$\langle a_k(t)a_{-k}(t) \rangle\, ; \langle
a^{\dagger}_k(t)a^{\dagger}_{-k}(t) \rangle$ which will be generated.
Now we can iterate the procedure obtaining the equivalent of eq.
(\ref{boltz}) but with the $N_k(t_0+\Delta t)$ on the right hand side,
iterating this procedure for all times is equivalent to the non-linear
equation
\begin{eqnarray}
{\dot{N}}_k(t) &=& {\lambda^2 \over 3} \, {1 \over 2 \, \omega_k} \,
\int {d^3 k_1 \over (2 \pi)^3} \, {d^3 k_2 \over (2 \pi)^3} \,
{d^3 k_3 \over (2 \pi)^3} \, {1 \over 8 \omega_{k1} \omega_{k2}
\omega_{k3} } \, (2 \pi)^3 \delta^3 ({\vec{k}}-{\vec{k}}_1 -
{\vec{k}}_2- {\vec{k}}_3)                                 \nonumber \\
&& \qquad\qquad\;
\times \big[ \delta( \omega_k + \omega_{k1} + \omega_{k2}+
\omega_{k3}) {\cal N}_1(t)                                \nonumber \\
&& \qquad\qquad\qquad\;
+ 3\, \delta( \omega_k + \omega_{k1} + \omega_{k2}- \omega_{k3})
{\cal N}_2(t)                                             \nonumber \\
&& \qquad\qquad\qquad\qquad\;
+ 3 \, \delta(\omega_k + \omega_{k1} -\omega_{k2}- \omega_{k3} )
{\cal N}_3(t)                                             \nonumber \\
&& \qquad\qquad\qquad\qquad\qquad\;
+ \delta( \omega_k -\omega_{k1} -\omega_{k2}- \omega_{k3}){\cal
N}_4(t)\big]
                                                        \label{boltzt}
\end{eqnarray}

Linearizing the above equation, considering only the mode with
wavevector $k$ to be slightly out of equilibrium whereas all the other
modes are in equilibrium leads to the ``relaxation time
approximation''
\begin{equation}
\frac{d \delta n_k}{dt} = -\frac{\Sigma_I(\omega_k)}{\omega_k} \delta
n_k(t)                                            \label{linearboltzt}
\end{equation}

The only term that gives a non-zero contribution in (\ref{boltzt}) is
the third (``scattering'') term, because for all the other terms the
$\delta$ functions have no support on the mass-shell. Therefore we
obtain the Boltzmann equation (\ref{boltzt}) and its ``relaxation time
approximation'' (\ref{linearboltzt}) (where only the second term in
(\ref{imaginary}) contributes to $\Sigma_I(\omega_k)$). This result
and the interpretation of the damping rate
${\Sigma_I(\omega_k)}/{2\omega_k}$ as (half) the relaxation rate of
the quasiparticle distribution function was proposed by
Weldon\cite{weldon}. To our knowledge it has not been proven before by
obtaining the Boltzmann equation directly from a first principle
calculation and analyzing the ``coarse graining'' approximations that
lead to this result.

The linearized approximation gives the time scales for relaxation for
situations close to equilibrium. In the case of small momentum
or $T>>M_R(T)>>p$ we obtain
\begin{eqnarray}
\tau_r(k \approx 0)&\approx&\frac{M_R(T)}{\Sigma_I(\omega_{k\approx
0})} \approx \frac{32 \sqrt{24\pi}}{\lambda^{3/2}T}  \label{trelax} \\
M_R(T)\tau_r &\propto & {1 \over \lambda} \gg 1      \label{timescale}
\end{eqnarray}
where we have assumed very weak coupling.

The collisional relaxation rate obtained from the hard thermal loop
resummation
\begin{equation}
\Gamma_{coll} = \frac{\lambda^{3/2}T}{32 \sqrt{24\pi}}
                                                 \label{collisionrate}
\end{equation}
is very {\em different} from the ``naive'' scattering rate that one
would write down for a ``classical'' Boltzmann equation. Such a
collision rate would be obtained as
\begin{equation}
\Gamma_{cla} = \langle n \sigma v \rangle              \label{naiverate}
\end{equation}
where $n$ is the number density, $\sigma$ the $2 \rightarrow 2$
scattering cross section, $v$ the mean velocity of the colliding particles
  and $\langle \cdots \rangle$ the ensemble
average. For $T \gg m$ one would obtain $n \approx T^3$, $v \approx 1$,
 the Born
cross section is $\sigma(E) \approx \lambda^2 / E^2$ where $E$ is the
total energy of the colliding particles, which in the high temperature
plasma is $E \approx T$. Such a ``classical'' estimate would lead to a
scattering rate $\Gamma_{cla} \approx \lambda^2 T$ which for very weak
coupling grossly underestimates the correct resummed result
(\ref{collisionrate}) for long wavelength modes.

\subsection{$T<T_c$}

In this section we obtain the Boltzmann equation in the broken
symmetry case. After a shift of the fields on both branches by the
vacuum expectation value as $\Phi^{\pm} \rightarrow v+\Phi^{\pm}$, the
interaction Hamiltonian contains new terms as compared to the unbroken
symmetry case of the previous section. These are given by
\begin{eqnarray}
H_{int}^v & = & \frac{\lambda v}{6}\left\{ \frac{1}{\sqrt{\Omega}}
\sum_{\vec k1, \vec k2, \vec k3} \Phi_{\vec k1} \Phi_{\vec k2}
\Phi_{\vec k3}\delta_{ \vec k1+ \vec k2 + \vec k3, \vec 0} + h
\sqrt{\Omega} \Phi_{\vec k=0}\right\}             \label{intTsmall} \\
h & = & v^2 - \frac{6 m^2_0}{\lambda}\ .                     \label{H}
\end{eqnarray}
There is now a one loop contribution to the Boltzmann equation. We
will {\em not} fix the v.e.v. to the tree level value. Rather, $h$ is
considered as a source term that will be used to cancel all the
tadpoles arising from the cubic interaction, whereas the mass
counterterm will cancel the tadpoles corresponding to mass
corrections. In this manner, using $h$ to cancel the cubic tadpoles
the result for $v$ is the {\em true} v.e.v. including quantum (and
thermal) corrections.

We now proceed in the same manner as in the unbroken symmetry case of
the previous section and obtain the evolution equation for the
occupation number by using the Heisenberg field equations and the
non-equilibrium Feynman rules. Using the canonical commutation
relations between the field and its canonical momentum, and requiring
that $h$ cancels all the tadpoles arising from the cubic interaction,
and $\delta m^2(T)$ cancel the tadpoles corresponding to mass
insertions, we find the new contribution to the evolution equation
\begin{eqnarray}
\dot{N}^{v}_k(t) & = & -\frac{\lambda v}{2 \omega_k}
\frac{\partial}{\partial t'} \left[ \langle (\Phi^+(t))^2_{k}
\Phi^-_{-k}(t')\rangle\right]_{t=t'}               \label{newboltz} \\
(\Phi^+(t))^2_{k} & = & \frac{1}{\sqrt{\Omega}} \sum_{\vec k1, \vec
k2} \Phi_{\vec k1}\Phi_{\vec k2} \delta_{\vec k1+\vec k2 \, ,\vec k}
                                                           \label{fi3}
\end{eqnarray}
This contribution must be added to the obtained in the previous
section and given by equation (\ref{ndotequa}). To one loop order, the
contributions are shown in figure 2 and to this order we obtain the
following evolution equation for the distribution functions:
\begin{eqnarray}
{\dot{N}}^v_k (t) &=& \frac{\lambda^2 v^2}{ 2 \, \omega_k} \, \int
{d^3 k_1 \over (2 \pi)^3} \, {d^3 k_2 \over (2 \pi)^3} \, {1 \over 4
\omega_{k1} \omega_{k2} } \, (2 \pi)^3 \delta^3 ({\vec{k}}-{\vec{k}}_1
- {\vec{k}}_2)                                            \nonumber \\
&&\qquad\;\times \Bigg\{
\left[\frac{ \sin \left[ ( \omega_k + \omega_{k1} + \omega_{k2})
(t-t_0) \right]} { \omega_k + \omega_{k1} + \omega_{k2}} \right]{\cal
N}_1^v(t_0)                                               \nonumber \\
&& \qquad\qquad\;
+ \left[\frac{ \sin \; \left[ ( \omega_k -\omega_{k1} -\omega_{k2})
(t-t_0) \right]} { \omega_k -\omega_{k1} -\omega_{k2}} \right]
{\cal N}_2^v(t_0)                                         \nonumber \\
&& \qquad\qquad\qquad\;
+ \left[\frac{\sin \left[ ( \omega_k + \omega_{k1}
-\omega_{k2})(t-t_0) \right]} {\omega_k + \omega_{k1} -\omega_{k2}
}\right] {\cal N}_3^v(t_0)                                \nonumber \\
&& \qquad\qquad\qquad\qquad\;
+ \left[ \frac{ \sin \; \left[ ( \omega_k - \omega_{k1} +
\omega_{k2} )(t-t_0) \right]} { \omega_k - \omega_{k1} + \omega_{k2} }
\right] {\cal N}_4^v(t_0)\Bigg\} \ ,                    \label{vboltz}
\end{eqnarray}
where
\begin{eqnarray}
{\cal N}_1^v(t) &=& (1+ N_k (t) )\,(1+ N_{k1} (t))\, (1+ N_{k2} (t))-
N_k (t) \, N_{k1} (t) \, N_{k2} (t)                       \nonumber \\
{\cal N}_2^v(t)&=& (1+ N_k (t) )\, N_{k1} (t) \, N_{k2} (t)- N_k (t)
\, (1+ N_{k1} (t)) \, (1+ N_{k2} (t))                     \nonumber \\
{\cal N}_3^v(t) &=& (1+ N_k (t) )\,(1+ N_{k1} (t)) \, N_{k2} (t)- N_k
(t) \, N_{k1} (t) \, (1+ N_{k2} (t))                      \nonumber \\
{\cal N}_4^v(t) &=& (1+ N_k (t) ) N_{k1} (t) \, (1+ N_{k2} (t))- N_k
(t) \,(1+ N_{k1} (t)) \, N_{k2} (t)                          \label{ocupanumber33}
\end{eqnarray}

Again under the assumption of a separation of time scales and
approximating the sine functions by energy conserving delta functions
using (\ref{goldenrule}) as before, and considering that all the modes
but the mode with wave-vector $k$ are in equilibrium with
distributions $N_{ki}(t_0)= n_{ki}$ and $N_k(t_0)=n_k+\delta n_k(t_0)$
we are led to the relaxation time approximation
\begin{equation}
\frac{d \delta n_k}{dt} = -\frac{\Sigma^v_I(\omega_k)}{\omega_k}
\delta n_k(t_0)                                  \label{linearboltztv}
\end{equation}
with $\Sigma^v_I(\omega_k)$ given by the imaginary part of the one
loop contribution to the self energy, equations (\ref{specdens}) and
(\ref{impar1lup}). In the broken symmetry case with only one scalar
field, the imaginary part $\Sigma^v_I(\omega_k)$ vanishes on-shell
because of kinematics. However, it is non-zero in the case in which
the particle $\Phi$ can decay into lighter scalars $\chi$ as discussed
in section III-B and below.

The discussion of the previous case also applies here, in that the
above evolution equations neglect the change of occupations numbers on
the right hand side of the equations, and are thus valid only for
early times. However, as discussed in the previous case, if there is a
separation of time scales such that $\tau_r M_R(T) >>1$ we can
integrate the evolution equations for times
$M^{-1}_R(T)<<(t-t_0)<<\tau_r$, approximate the sine functions by energy
conserving delta functions, and keep the occupation numbers fixed at
their values at $t_0$. Furthermore reducing the density matrix to the
diagonal elements in the basis of occupation numbers at time $t$, we
can iterate this equation leading to the usual Boltzmann equation. The
new terms describe the evolution of the distribution function arising
from processes of decay and recombination in the plasma. Such an
iterative procedure leads to the following one-loop contribution to
the Boltzmann equation
\begin{eqnarray}
{\dot{N}}^v_k (t) &=& \frac{\lambda^2 v^2}{ 2 \; \omega_k} \; \int
{d^3 k_1 \over (2 \pi)^3} \, {d^3 k_2 \over (2 \pi)^3} \, {1 \over 4
\omega_{k1} \omega_{k2} } \, (2 \pi)^3 \delta^3 ({\vec{k}}-{\vec{k}}_1
- {\vec{k}}_2)                                            \nonumber \\
&& \qquad\times \delta ( \omega_k + \omega_{k1} + \omega_{k2}) {\cal
N}_1^v(t)                                                 \nonumber \\
&& \qquad\qquad + \delta ( \omega_k  -\omega_{k1}  -\omega_{k2})
{\cal N}_2^v(t)                                           \nonumber \\
&& \qquad\qquad\qquad + \delta ( \omega_k + \omega_{k1}
-\omega_{k2}){\cal N}_3^v(t)                              \nonumber \\
&& \qquad\qquad\qquad\qquad
+ \delta ( \omega_k - \omega_{k1} + \omega_{k2} )
{\cal N}_4^v(t)\ .                                     \label{vboltzt}
\end{eqnarray}

This one-loop result also applies if the scalar field $\Phi$ couples
trilinearly to lighter scalars $\chi$, namely with a vertex $g \Phi
\chi^2$ and we assume that kinematical decay is possible, i.e.
$M_R(T)> 2 m_{\chi}(T)$. Under the assumption of separation of time
scales, which is the case for small coupling, justifying the
replacement of the sine by delta functions we obtain a Boltzmann
equation for evolution of the distribution function for the $\Phi$
field which has the same form as (\ref{vboltzt}), but with
$\lambda^2v^2/2$ replaced by $2g^2$ and the functions ${\cal
N}_i^v(t)$ replaced with
\begin{eqnarray}
{\cal N}_1^{\chi}(t)&=& (1+ N_k (t))\,(1+ N_{k1}^{\chi}(t))\, (1+
N_{k2}^{\chi}(t))- N_k (t) \,N_{k1}^{\chi}(t) \, N_{k2}^{\chi}(t)
                                                          \nonumber \\
{\cal N}_2^{\chi}(t)&=& (1+ N_k (t) )\, N_{k1}^{\chi} (t) \,
N_{k2}^{\chi} (t)- N_k (t) \, (1+ N_{k1}^{\chi}(t)) \, (1+
N_{k2}^{\chi}(t))                                         \nonumber \\
{\cal N}_3^{\chi}(t)&=& (1+ N_k (t))\,(1+ N_{k1}^{\chi}(t)) \,
N_{k2}^{\chi}(t)- N_k (t) \, N_{k1}^{\chi} (t) \, (1+ N_{k2}^{\chi}
(t))                                                      \nonumber \\
{\cal N}_4^{\chi}(t)&=& (1+ N_k (t) ) N_{k1}^{\chi} (t) \, (1+
N_{k2}^{\chi} (t))- N_k (t) \,(1+ N_{k1}^{\chi}(t)) \, N_{k2}^{\chi}
(t)\ .                                                       \label{ocupanumber44}
\end{eqnarray}
Here $N_k(t)$ and $N_k^{\chi}(t)$ are the non-equilibrium distribution
functions for the quanta of the fields $\Phi$ and $\chi$ respectively
and $\omega_{k1,2}= \sqrt{k^2_{1,2}+m^2_{\chi}(T)}$. The term
proportional to ${\cal N}_2^{\chi} (t)$ describes the decay process of
$\Phi \rightarrow \chi \chi$ minus the ``recombination'' $\chi \chi
\rightarrow \Phi$ in the medium. The relaxation time approximation for
the evolution of $N_k(t)=n_k+\delta n_k(t)$ while the $N_k^{\chi}$ are
the equilibrium functions is given by
\begin{equation}
\frac{d \delta n_k}{dt} = - 2 \Gamma_{\Phi \leftrightarrow \chi \chi}
\delta n_k(t)                                       \label{reltimeapp}
\end{equation}
with $ \Gamma_{\Phi \leftrightarrow \chi \chi}$ the decay minus
recombination rate given by equation (\ref{decayrate}).

Thus, after this analysis the nature and validity of the
approximations and resummations implied in the usual Boltzmann
equation (and its linearized version) become clear:

{\bf i)} For high temperatures a hard thermal loop resummation to
incorporate the proper microscopic time scales. Such a resummation is
needed to incorporate the relevant time for ``coarse graining''.

{\bf ii)} A wide separation between the microscopic time scale and
the relaxation time which, as shown by the estimate (\ref{timescale}),
is warranted for very weak coupling. Such a case corresponds to the
collisional and decay lifetime of the particle being much larger than
the microscopic time scale given by the inverse mass of the particle
in the medium, namely a ``sharp resonance'' condition. This separation
of time scales permits the identification of intermediate times
$M^{-1}_R(T) \ll t \ll \tau_r$ such that the ``Fermi Golden Rule''
approximation (\ref{goldenrule}) is justified and energy conservation
is enforced.

{\bf iii)} A reduction of the density matrix to the diagonal elements
in terms of the time dependent occupation number. This reduction
neglects higher order correlations of the quasiparticle operators such
as $\langle a_k(t) a_k(t) \rangle$ and $\langle a^{\dagger}_k(t)
a^{\dagger}_k(t) \rangle$ . Therefore the ``coarse graining''
procedure is not only at the level of spatial and temporal averages
but also of reducing the density matrix, thus reducing the hierarchy
of equations of motion.

Although all of these approximations are implicit in the usual
procedure through the Wigner transform and the gradient expansion, our
method illuminates all the necessary steps and the physical
justifications at once.

\subsection{Improvements:}

This method allows us to identify several improvements to the
description the kinetics of thermalization and relaxation:

{\bf i)} The first improvement is to incorporate wave-function
renormalization in the definition of the quasiparticles in the
Hamiltonian as well as coupling constant renormalization. For this we
write the renormalized fields, canonical momentum and coupling
constant as
\begin{equation}
\Phi_R = Z^{-1/2}_{\Phi} \Phi \, ; \; \; \Pi_R = Z^{1/2}_{\Phi} \Pi \,
; \; \; \lambda_R = Z^2_{\Phi}Z^{-1}_{\lambda} \lambda
                                                      \label{wavefunc}
\end{equation}
Notice that $\Pi_R$ and $\Phi_R$ are canonical conjugate variables.
The ``free quasiparticle'' Hamiltonian is now written in terms of
$\Pi_R$, $\Phi_R$ and $M_R(T)$ and the difference between the bare and
renormalized quantities is accounted for in the interaction part of
the Hamiltonian. The quasiparticle distribution $N_k(t)$ is defined as
in eq. (\ref{ocupation}) but in terms of the renormalized fields and
canonical momenta. Its evolution equation is obtained from the
Heisenberg equations of motion with the new terms in the interaction
Hamiltonian as described previously.

There will now be new terms in the equations for the evolution of the
distribution function that will cancel the (divergent) contributions
arising from the wave function and coupling constant renormalization.
These terms will appear at three loops (for coupling constant) and
beyond but in terms of the quasiparticle propagators. This is
precisely the BPHZ program of renormalization but here implemented in
a non-equilibrium formulation. Clearly such an improvement comes at
the cost of a complicated structure but this method allows us to
incorporate such corrections in a systematic manner.

{\bf ii)} Rather than assuming a wide separation of time scales and
using the approximation (\ref{goldenrule}) leading to energy
conserving delta functions, one can integrate the evolution equations
(\ref{boltz}, \ref{vboltz}) numerically in increments $t=t_0+\Delta
t$, ``reseting'' the values of the occupation numbers for the next
increment to those obtained in the previous iteration. Such an
approximation can be implemented numerically just as in the usual
case. However the equation for the evolution of the distribution
functions will now have {\em off-shell} contributions that will modify
the evolution on time scales comparable to $M^{-1}_R(T)$. These
contributions include information on the preparation of the
non-equilibrium state and make the derivative of the occupation number
vanish at $t=t_0$ as it must (the same feature appears in the
derivation of Fermi's Golden Rule).

In the two-loop collisional contribution, the energy-conserving delta
functions introduce one constraint. Thus a five dimensional k-integral
remains, whereas in the ``improved'' form, without energy conservation
one must face the full six-dimensional k-integral. Furthermore, this
``improved'' set of evolution equations include terms that change
chemical equilibrium (do not conserve particle number) on microscopic
time scales. These result from the fact that the interactions
considered in this study do not conserve particle number and may lead
to new effects in the quasiparticle distribution at long times. Such a
possibility should be studied further.

\section{Conclusions and Implications}

In this article we have presented a new approach, based on
non-equilibrium quantum field theory, to study linear relaxation of
quasiparticles and the kinetic description of the evolution of the
quasiparticle distribution functions in a scalar field theory. Our
method incorporates the relevant microscopic time scales through the
resummation of hard thermal loops in the description of both
relaxation and kinetics. The method leads to a systematic and
consistent scheme for obtaining both the linearized equations of
motion and the equations for the distribution functions. It clearly
displays the different ``coarse graining'' approximations usually
involved in a kinetic description and their range of validity. It
allows for a systematic and consistent improvement in the kinetic
description, including off-shell and renormalization effects, as well
as processes that change chemical equilibrium on short time scales,
and is numerically implementable. These improvements in the kinetic
equations may prove physically relevant if the time scale for
relaxational and thermalization processes is not widely different from
the microscopic time scales and off-shell processes must be taken into
account for a more accurate description.

As a result of our study of both relaxation and kinetics we
established a direct proof of the relationship first proposed by
Weldon\cite{weldon} between the damping rate of the quasiparticles and
the relaxation rate of the quasiparticle distribution function in the
relaxation time approximation after the coarse graining of the kinetic
equations.

We have studied relaxation and kinetics both in the unbroken symmetry
phase for $T>>T_c$ as well as in the ordered, broken symmetry phase
for $T<T_c$. The hard thermal loop resummation is very important in
the unbroken phase to obtain the proper microscopic time scales and to
establish a wide separation of time scales that justifies a kinetic
description in the weak coupling regime. The broken symmetry case
provides a setting to incorporate processes in which the scalar
particle can decay, via a trilinear coupling, to other lighter scalars
in the theory. We studied the effect of such processes both on
relaxation and kinetics, obtaining the kinetic equations that include
decay and recombination in the medium. We also analyzed the effect of
processes that only contribute below the light cone that are similar
to Landau damping. We pointed out that although these processes do not
contribute {\em directly} to relaxation and damping rates, they
contribute {\em indirectly} through the sum rules obeyed by the
spectral function.

The methods developed here are suitable for treating the evolution of
the distribution functions for the particles produced during the
parametric amplification stage in inflationary cosmological
scenarios\cite{kofman,branden3,reheating,yoshimura} and in the
supercooling stage of the quark-gluon transition \cite{muller2}. The
reason is that these methods can be simply extended to include the
time evolution of the expectation value of the scalar fields (mean
fields) into the kinetic equations. If there is a separation of time
scales between the stage of particle production and the onset of
thermalization, these equations will lead to a description of particle
production via parametric amplification {\em and} the collisional
processes leading to thermalization and completion of the reheating
stage. We are currently applying these techniques to that problem, and
we expect that these methods will be extended to gauge and fermion
theories in the near future.

Some questions that this study poses that are important are the
following: is there a kinetic description of ``critical slowing down''
near a phase transition? can a kinetic description be extended to
temperatures near criticality when long range correlations begin to
form and relaxational time scales should diverge?

These questions are relevant for a deeper understanding of the
dynamics of phase transitions and the non-equilibrium description of
thermalization and relaxation of long-wavelength fluctuations. We
expect to address these and other related questions in the near
future.

\acknowledgements

D.-S. Lee would like to thank J. Ng, B.-L. Hu and L. Dolan for
interesting discussions. D.B. would like to thank H. J. de Vega, M.
D'Attanasio, R. Holman, R. Pisarski, H. A. Weldon, P. Elmfors, P.
Henning and J. L. Goity for illuminating conversations during the
early stages of this work. He also thanks the N.S.F. for partial
funding through grant award: PHY-9302534. D.-S. Lee was supported in
part by DOE grant DE-FG05-85ER-40219-TASKA. I.D.L. thanks the U.K.
Particle Physics and Astronomy Research Council for a travel grant.

\newpage

{\underline{Figure Captions:}}

{\bf Fig. 1} Top: two-loop diagrams contributing to the equation of
motion, the dashed line represents an insertion of the non-equilibrium
expectation value. Bottom: two-loop diagrams contributing to the
Boltzmann equation.

{\bf Fig. 2} Top: one-loop diagrams contributing to the equation of
motion in the broken symmetry phase. Tadpoles have been absorbed in a
mass and v.e.v. renormalization. The dashed line represents an
insertion of the non-equilibrium expectation value. Bottom: one loop
diagrams contributing to the Boltzmann equation in the broken symmetry
phase.

\end{document}